# A Review on Elliptic Curve Cryptography for Embedded Systems

Rahat Afreen[1] and S.C. Mehrotra[2]

[1]Tom Patrick Institute of Computer & I.T, Dr. Rafiq Zakaria Campus, Rauza Bagh, Aurangabad. (Maharashtra) INDIA
`siddiqui.rahat@gmail.com`

[2]Department of C.S. & I.T., Dr. B.A.M. University, Aurangabad. (Maharashtra) INDIA
`mehrotra_suresh@yahoo.com`

## ABSTRACT

*Importance of Elliptic Curves in Cryptography was independently proposed by Neal Koblitz and Victor Miller in 1985. Since then, Elliptic curve cryptography or ECC has evolved as a vast field for public key cryptography (PKC) systems. In PKC system, we use separate keys to encode and decode the data. Since one of the keys is distributed publicly in PKC systems, the strength of security depends on large key size. The mathematical problems of prime factorization and discrete logarithm are previously used in PKC systems. ECC has proved to provide same level of security with relatively small key sizes. The research in the field of ECC is mostly focused on its implementation on application specific systems. Such systems have restricted resources like storage, processing speed and domain specific CPU architecture.*

## 1. INTRODUCTION

The changing global scenario shows an elegant merging of computing and communication in such a way that computers with wired communication are being rapidly replaced to smaller handheld embedded computers using wireless communication in almost every field. This has increased data privacy and security requirements. Data protection and authentication is now demanded for performing mobile banking on a cell phone, monitoring health of a patient through his wrist watch, remaining connected to office networks while travelling and so on. This has given a new thrust of what the technology guru Eddie Murphy has called the fourth wave – "Universal Connectivity" also termed as "Communication and Connectivity" by Embedded Market Forecasters. First three waves being defined as mainframe computers, PC revolution and Internet explosion respectively [1]. Data and information security is equally required along with other basic needs of reliable connectivity; high data transfer rate, optimised storage and processing etc.

Another scenario to consider is that although we are defining embedded systems to be highly domain specific but, one the other side domains themselves are expanding for such systems. Consider the example of microwave oven- the only purpose of this device is to provide temperature and timing control. This can be achieved from a basic program which can be implemented directly using machine language on any micro-controller. And now, we have ovens in the market with pre-programmed temperature and time settings for common food items and recipes. Further, SB electronics systems, UK have launched an oven in their famous iWave range which can re-order the stock that is used in a restaurant [2]. It is equipped with a barcode reader to identify packed food, can be connected to the inventory database via GPRS,

International Journal of Computer Science & Information Technology (IJCSIT), Vol 3, No 3, June 2011

modem or infrared connection for a laptop and even if that is not enough it can have its own identifier to recognize individual oven if they are in multiple. Thus, manufacturers are planning to connect to their devices to offer better service; every printer, elevator, air conditioner, vending machine, etc., can report its status, financial receipts and maintenance requirements as they occur. In short functional requirement of even such basic electronic gadgets are increasing, resulting into the requirement of more comprehensive software development platforms. This has resulted into the introduction of embedded operating systems and compilers of various high level languages for embedded systems. The progress is almost on the similar lines how computer systems have evolved into various layers of hardware, operating systems and application programs, which later got clubbed with communication networks. Similarly now embedded devices are also getting connected for information transfer and hence the need of network security is arising for these domain specific systems. As these systems are classified to be resource constrained, the small key size of ECC makes it effective to implement on such systems. Following table shows a comparison of elliptic curve keys with Diffie Hellman Keys[3].

Table 1. Strength of Diffie Hellman V/s Elliptic Curve Keys.

| Security Level (bits) | Ratio of DH Security: EC Security |
|---|---|
| 80 | 3:1 |
| 112 | 6:1 |
| 128 | 10:1 |
| 192 | 32:1 |
| 256 | 64:1 |

ECC is emerging as a most trusted solution for providing security on embedded systems.

Section II of this review describes the fundamental concepts of modular arithmetic that forms the basis for today's encryption systems. Section III defines elliptic curves from various aspects and explains its basic point addition and point doubling operations. Section IV is dedicated to the overview of ECC process as a whole while Section V gives some of the algorithms used for ECC performance optimisation. As finite field is the equally important aspect for designing cryptographic systems, Section VI explains various possible finite fields that can be considered while implementing ECC. Accordingly, this section also introduces ECC domain parameters and ECC protocol algorithms. Various design considerations available currently for ECC implementation are discussed in section VII along with available ECC standards, followed by conclusion in section VIII.

## 2. MODULAR ARITHMETIC

Before proceeding to a detailed description of Elliptic Curves, let us revise basics of Modular Arithmetic as this aspect of mathematics is very closely related to today's cryptographic horizon and form the base for Elliptic Curve Cryptography.

Modular arithmetic is a branch of number theory, which allows reformulation of the way addition, subtraction and multiplications are performed [4]. It is related to Number Theory that can be defined as, "The study of integers". Number theory is one of the oldest and largest branches of pure mathematics. It is also called as Higher Arithmetic. It consists of study of the properties of whole numbers, primes, prime factorisation etc. basically number theory is full of questions that are easy to pose but difficult to answer. Computer scientists use this property for

85

their advantage. To sum up, Gauss called mathematics as the queen of sciences and considered number theory as the queen of mathematics.

Modular arithmetic is arithmetic operations in the Finite Field or Galois field GF(n) named in the honour of Évariste Galois. It has several applications in coding theory, computer algebra, and cryptography. Efficient modular arithmetic algorithms play an important role in today's cryptographic systems. Most practical public key systems exploit the properties of arithmetic in large finite groups. For methods such as Diffie–Hellman and Elliptic Curve cryptosystems, security depends on the contrast in difficulty between performing two group operations: exponentiation vs. discrete logarithm. The discrete log problem is believed to be hard compared to the exponentiation problem; and the elliptic curve discrete logarithm problem is even harder. This is because of its different algebraic structure, it's complex arithmetic rules to "add" two points on an elliptic curve, and the lack of an index calculus method for the elliptic curve domain.

### 2.1. Modular Multiplication

Modular multiplication is simply the computation of the remainder of the product of two numbers with respect to a modulus. More formally, the modular multiplication problem is defined as the computation of R = A × B mod M given the integers A, B, M with $0 \leq A, B < M$. [5]

### 2.2. Irreducible Polynomial

Irreducible polynomial is an analogue to modulus p in modular arithmetic. Irreducible polynomial is a polynomial of degree m that cannot be expressed as the product of two polynomials of lesser degree. In many standard implementation of elliptic curve operation, for making polynomial reduction more efficient the irreducible polynomial is chosen to be trinomial or pentanomial.

## 3. ELLIPTIC CURVE DEFINITION

The generalized equation of elliptic curve E over a field K is given by

$$E: y^2 + a_1 xy + a_3 y = x^3 + a_2 x^2 + a_4 x + a_6$$

This equation is called a Weierstrass equation. Where a1, a2, a3, a4, a6 ∈ K and $\Delta \neq 0$. The $\Delta$ is the discriminant of E and is defined as follows:

$$\Delta = -d_2^2 d_8 - 8 d_4^3 - 27 d_6^2 + 9 d_2 d_4 d_6$$

Where

$$d_2 = a_1^2 + 4 a_2$$
$$d_4 = 2 a_4 + a_1 a_3$$
$$d_6 = a_3^2 + 4 a_6$$
$$d_8 = a_1^2 a_6 + 4 a_2 a_6 - a_1 a_3 a_4 + a_2 a_3^2 - a_4^2 \ldots\ldots\ldots \quad (1.2)$$

E is defined over K because the coefficients $a_1, a_2, a_3, a_4, a_6$ of its defining equation are elements of K. E(K) emphasize that E is defined over K, and K is called the underlying field.

Elliptic curves are not like an ellipse or curve in shape. They look similar to doughnuts. Geometrically speaking they somehow resemble the shape of torus, which is the product of two circles when projected in three-dimensional coordinates[6].

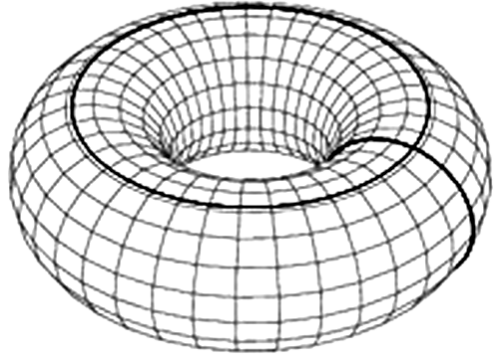

Figure 1. A Torus

Elliptic curves can more generally be defined over any finite field. In particular, the characteristic two finite fields F(2m) and Fp are of special interest since they lead to the most efficient implementation of the elliptic curve arithmetic.

The elements of Fp are integers in the range [0, 1, 2, …p-1] wehre p is a prime. The simplified form of Weierstrass equation for Fp is

$$y^2 = x^3 + ax + b (\text{mod } p) \qquad (1.3)$$

For example, the points for P=29 in E (F29) are

∞   (2,6)  (4,19)  (8,10)  (13,23) (16,2)  (19,16)  (27,2)

(0,7)  (2,23)  (5,7)  (8,19)  (14,6)  (16,27)  (20,3)  (27,27)

(0,22) (3,1)  (5,22)  (10,4)  (14,23) (17,10) (20,26)

(1,5)  (3,28)  (6,12)  (10,25)  (15,2)  (17,19)  (24,7)

(1,24) (4,10)  (6,17)  (13,6)  (15,27) (19,13) (24,22)

Examples of elliptic curve addition are (5,22) + (16,27) = (13,6), and 2(5,22) =(14,6) Using point addition and point doubling rules of elliptic curve. The elements of F(2m) are simply binary numbers but with (m-1) as highest power of 2, m being prime. But these numbers are considered as polynomials of the form

$$a_{m-1}z^{m-1} + a_{m-2}z^{m-2} + ... + a_2z^2 + a_1z + a_0$$

where a∈ {0,1}.These numbers are called as binary polynomials and arithmetic operations on them are redefined accordingly.[7]

### 3.1. Geometrical Definition of Point Addition and point Doubling

For any two points P(x1,y1) ≠ Q(x2,y2) on an elliptic curve, EC group law point addition can be defined geometrically as: "If we draw a line through P and Q, this line will intersect the elliptic curve at a third point(-R). The reflection of this point about x-axis, R(x3,y3) is the addition of P and Q."

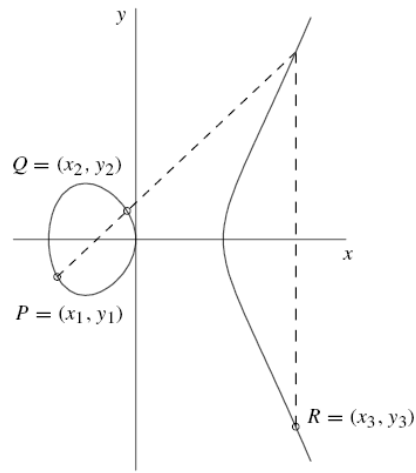

Figure 2. Addition: R=P+Q

It is calculated using following equations:

$$x_3 = \lambda^2 - x_1 - x_2$$

$$y_3 = \lambda(x_1 - x_3) - y_1$$

and $\lambda = \dfrac{y_2 - y_1}{x_2 - x_1}$  if $P \neq Q$

For P=Q , point doubling, Geometrically if we draw a tangent line at point P, this line intersects elliptic curve at point a point (-R). Then, R is the reflection of this point about x-axis.

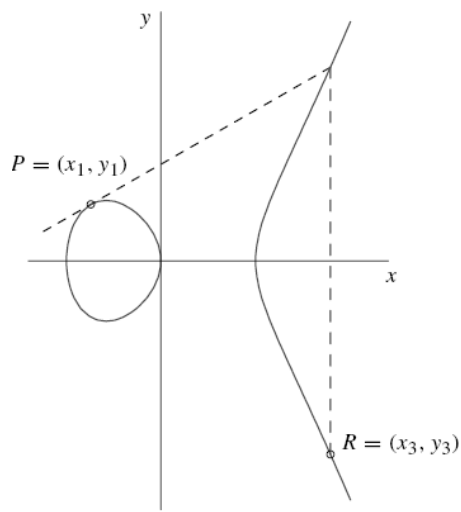

Figure 3. Doubling: R=P+P

Point doubling is calculated using following equations:

$$2P = (x_3, y_3), \text{ where}$$

$$x_3 = \lambda^2 - 2x_1 \text{ and}$$

$$y_3 = \lambda(x_1 - x_3) - y_1$$

where $$\lambda = \frac{3x_1^2 + a}{2y_1}$$ if P=Q

## 4. ELLIPTIC CURVE CRYPTOGRAPHY

The security of Elliptic Curve Cryptography relies on the difficulty of solving the Elliptic Curve Discrete Logarithm Problem ECDLP, which states that, "Given an elliptic curve E defined over a finite field Fq , a point P ∈ E(Fq ) of order n, and a point Q ∈ E(Fq ), find the integer k ∈ [0,n −1] such that Q = k P. The integer k is called the discrete logarithm of Q to the base P, denoted k = logP Q."

This point multiplication is performed by repeated point addition and point doubling for example 7P=(2((2P)+ P)+P. k is used as a private key and curve's base point G is used as a public key. For ECC we only consider those points which lie in some finite field. Q=kP is called scalar point multiplication. This is most time consuming operation in ECC implementation. The hierarchy of EC arithmetic is given in Fig. 4

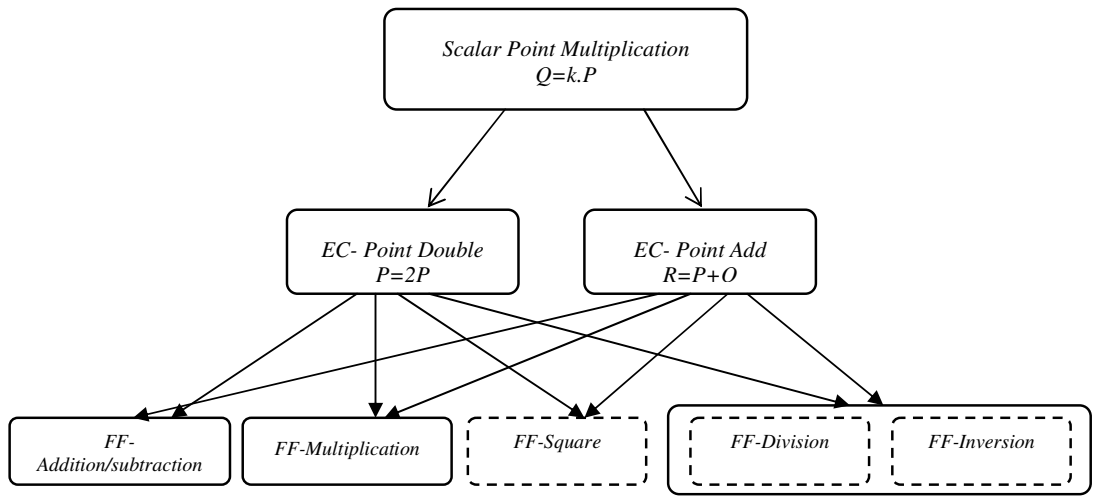

Figure 4  EC Arithmetic Architecture

The top level k.P algorithm is performed by repeated EC-Add and EC-Double operations. The EC operations in turn are composed of basic operations in the underlying finite field (FF) like FF-Addition/subtraction, FF-Multiplication. FF–square can be implemented using multiplication and as we are considering resource constrained devices this is generally preferred. The FF inversion is a very expensive operation, it is used to implement FF-Division Euclid's method is a common choice to implement inversion in finite field. But, Using

Montgomery's method and a special case of Euclid's inversion. FF- Division can be implemented directly [8]. The elliptic curve point multiplication Q=kP can be performed according to a variety of approaches. They are summarized in [9], as shown in Fig.5.The scalar k can have different representations. There is verity of algorithms to perform multiplication. Also there are various combinations of finite filed representation and coordinate system for curve points. A multitude of algorithms is available to be applied for each task. A proper combination of various algorithms can significantly affect the performance of EC point multiplication. In general, affine and projective coordinate systems are considered to perform calculations.

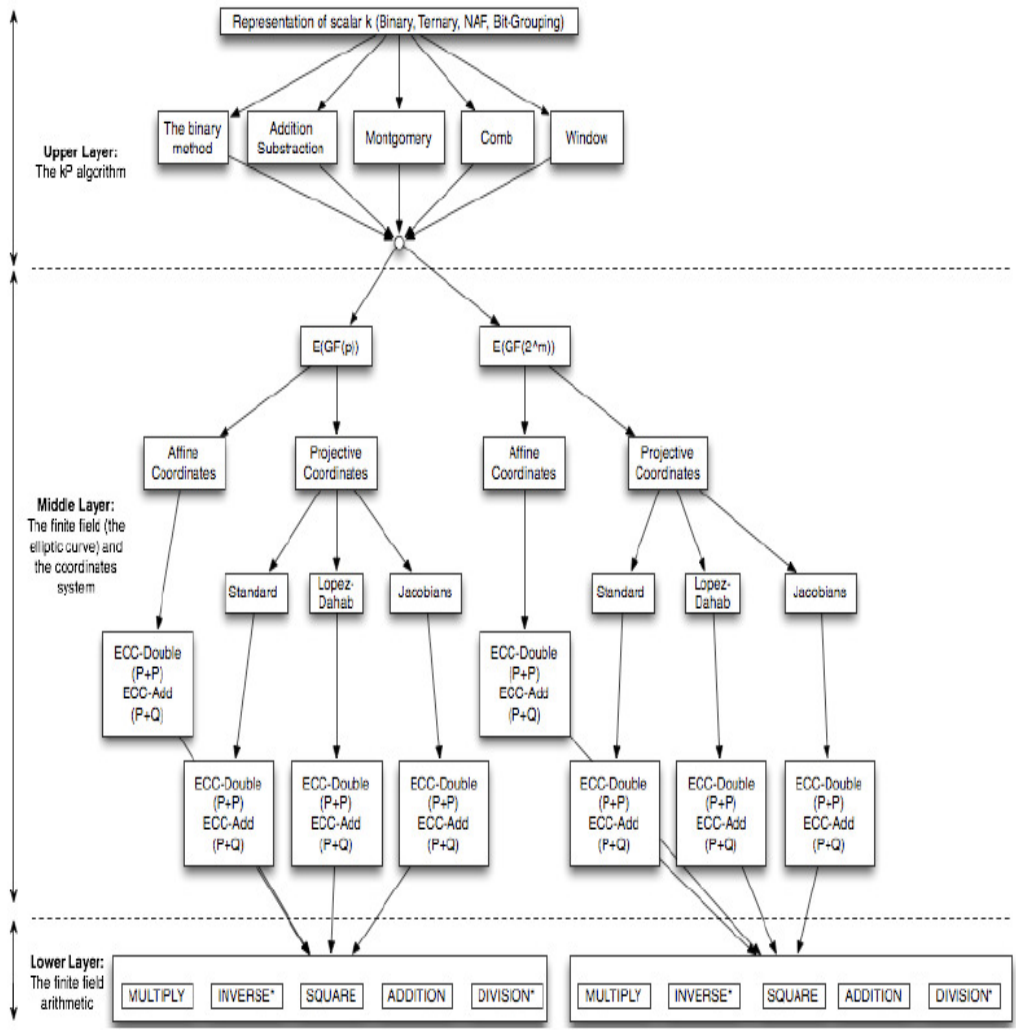

Figure 5. Various methods of scalar point multiplication k.P

Affine coordinates have the property that a set of n vectors forms the axes of this system and there is no clear definition of origin. To get an idea of the affine space, consider a bucket full of vectors. Each vector could be represented by an ordered pair of real numbers called the coordinates of the vectors. This could be represented as R2 .Now, consider an infinitely large square E2. So, the E2 will be affine space, consisting of any two non parallel vectors from R2 and a point from E2 as origin. Vectors in R2 are just floating around all the time, they have no fixed position. But points in E2 have fixed positions. [10][11][12]

The ECC arithmetic operations explained previously are with reference to affine space. The equation of elliptic curve in projective space (an affine space with Z plane )is given by:

$$Y^2Z = X + a_4XZ^2 + a_6Z^3$$

The point (X1 : Y1 : Z1) on E corresponds to the affine point (X1/Z1, Y1/Z1) when Z1 $\neq$ 0 and to the point at infinity P$\infty$ = (0 : 1 : 0) otherwise. The opposite of (X1 : Y1 : Z1) is (X1 : −Y1 : Z1). Arithmetic operations in projective space can be referred from [11].

## 5. ALGORITHMS FOR ECC PERFORMANCE OPTIMIZATION

Various algorithms to optimise the performance of elliptic curve multiplication, squaring, point multiplication etc are in use today some of the famous algorithms are given below;

### 5.1. Karatsuba Multiplication

Using a method developed by Karatsuba and Ofman, the number of multiplications for a polynomial equation can be reduced in ex-change for an increased number of additions. As long as the time ratio for executing a multiplication vs. an addition is high, this tradeoff is more efficient.

Consider the example of two degree-1 polynomials, $A(x) = a_1x + a_0$ and $B(x) = b_1x + b_0$. For the traditional method, we must calculate the product of each possible pair of coefficients. $D_0 = a_0b_0$, $D_1 = a_0b_1$, $D_2 = a_1b_0$, $D_3 = a_1b_1$. And then the product :C(x) = A(x) . B(x) is:

$$C(x) = D_3x^2 + (D_2 + D_1)x + D_0$$

The Karatsuba method begins by taking the same two polynomials, and calculating the three products

$$E_0 = a_0b_0, \quad E_1 = a_1b_1, \quad E_2 = (a_0 + a_1)(b_0 + b_1)$$

These are then used to assemble the result

C(x) = A(x) . B(x);

$$C(x) = E_1x^2 + (E_2 - E_1 - E_2)x + E_0$$

It is easy to verify that results are equal.

The traditional method requires four multiplications and one addition, while the Karatsuba method requires three multiplications and four additions. Thus karatsuba method has a single multiplication for three additions. If the cost to multiply on the target platform is as least three times the cost to add, then the method is effective. While this basic form of Karatsuba was presented in the original paper, there are a number of ways this method may be expanded to handle larger degree polynomials. This is shown in [16], where the authors give an in-depth study of this method and its variations.

### 5.2. Itoh-Tsujii Inversion

Extension field inversion is normally a costly operation, but the nature of OEFs (Optimal Extension Fields, Explained below;)allow the reduction of the extension field inversion to a subfield inversion. The Itoh-Tsujii algorithm which was originally developed for use with composite fields $GF(2^{n^m})$ in a normal basis representation can be applied to extension fields $GF(q^m)$ in polynomial representation. It is assumed that the subfield inverse can be calculated by efficient means, such as table-lookup or the Euclidean algorithm, given a small order of the subfield. To perform the OEF inversion, we use the following expression:

$A^{-1} = (A^r)^{-1} A^{r-1}$, Where $r = \dfrac{q^m - 1}{q - 1}$ .This equation shows the general case for inversion.

### 5.3. de Rooij Point Multiplication

for Q=k.P, well-studied techniques used for ordinary integer exponentiation can be advantageously adapted. The most basic of these algorithms is the binary-double-and-add algorithm It has a complexity of log2(k) + WH(k) group operations, where WH is the Hamming weight of the multiplier k. On average, we can expect this algorithm to require 1.5 log2(k) group operations. Using a method devised by de Rooij in [15], we are able to reduce the number of group operations necessary by a factor of four over the binary-double-and-add algorithm.

### 5.4. The Montgomery Modular Multiplication Algorithm

This is a very famous algorithm for modular multiplication. Its multiple implementations are available both in hardware and software as it is capable to speed up the modular multiplication process by five times [16]. The basic idea behind Montgomery multiplication is the fact that one can add a multiple of the modulus M to the product A·B to yield a result that is at most 2n+1 bits wide. Adding, instead of subtracting, a multiple of the modulus M does not affect the computation, since the result will be congruent to A·B modulo M. Two numbers are said to be congruent if their remainder when divided by the modulus is the same. Thus, A·B, A·B +M, A·B +2M ... A·B + kM are all congruent modulo M. This implies: A·B ≡ A·B +M ≡ A·B +2M ≡ ... (A·B + kM) mod M. In the Montgomery algorithm, the multiple of the modulus M that is added to A·B is chosen in such a way that the lower n-bits of the 2n+1-bit result are all zeroes [17]. The least significant half of the 2n+1-bit result that are all zeroes is then discarded. This way, the result would have been reduced to at most n+ 1 bit in width. A single subtraction of the modulus M can then be performed to further reduce the result to at most n- bits and make it less than M if required. It has been shown by Walter [18] that the extra subtraction may not be necessary under certain conditions.

It's easy to explain in decimal, but Montgomery multiplication is easier to implement in binary [19]. The place of 10n is taken by some suitable power of 2, but the key simplification is that the adding of the multiple of the modulus becomes much easier. The rule is this: if the number you are looking at is odd (a 1 in LSB), add R before you halve it; if it's even(a 0 in LSB), just halve it. Halving a number in binary is simply discarding its lower significant bit. Eg: (1100 is binary 12 if we discard LSB zero, we get 110, which is binary 6). Binary shift and rotate instructions are available in every machine level instruction set and can be used for this purpose. A very simple form of hardware based Montgomery Modular Multiplication algorithm is given in [17].

## 6. CHOICE OF FINITE FIELDS

For performing ECC we also need to select any one suitable finite field GF(). Various finite fields admit the use of different algorithms for arithmetic. This choice can have a dramatic impact on the performance of the ECC. In particular, there are generic algorithms for arithmetic in an arbitrary finite field and there are specialized algorithms, which provide better performance in finite fields of a particular form. Below is the summary of field types proposed for ECC.

### 6.1. Binary Fields

Implementers designing custom hardware for an ECC often choose p = 2 and P(x) to be a trinomial or pentanomial. Such choices of irreducible polynomial lead to efficient methods for extension field modular reduction. This type of fields is called binary fields. The elements of the subfield GF(2) can be represented by the logical signals 0 and 1. In this way, it is possible to construct fast and area efficient hardware circuits to perform the finite field arithmetic. Binary fields are also popular for software implementations of ECC.

### 6.2. Binary Composite Fields

In software, the choice of parameters varies considerably with the wide array of available microprocessors. Many authors have suggested the use of p = 2 and m a composite number, e.g. [20], [21]. In this case, the field GF(2m) is isomorphic to GF((2s)r), for m = sr and we call this a "composite field." Then multiplication and inversion in the subfield GF(2s) can be efficiently performed by index table look-up if s is not too large. In turn, these operations in the extension field GF((2s)r) are computed using arithmetic in the subfield. As in the binary field case, the irreducible polynomials for both the subfield and the extension field are chosen to have minimal weight. This approach can provide superior performance when compared to the case of binary fields. However, a recent attack against ECCs over composite fields [22] makes their use in practice questionable.

### 6.3. Prime Fields

Prime fields are perhaps the most obvious finite fields to use. For ECC, a typical prime is chosen to be larger than 2160, and must be stored in multiple computer words. The problem with this representation is that during computation, the carries between words must be propagated, and the reduction modulo p must be performed over several words. There has been a large amount of research dealing with methods for doing long-number multiprecision arithmetic efficiently. The most popular method in this context is based on Montgomery reduction explained previously.

### 6.4. Optimal Extension Fields

An alternative construction is to use optimal extension fields (OEFs), defined as follows. Choose a prime p of the form $2^n \pm c$, for n, c arbitrary positive integers, where $\log_2 c \leq \lfloor \frac{1}{2} n \rfloor$ In this case, one chooses p of appropriate size to use the multiply instructions available on the target microcontroller. In addition, m is chosen so that an irreducible binomial $P(x) = x^m - \omega$ exists, $\omega \in GF(p)$. To generate "good" elliptic curves over OEFs there are two basic approaches. The first one is based on the use of a curve defined over GF(p). The second, more general, method uses Schoof's algorithm together with its improvements. The field is chosen with finitely large number of points suited for cryptographic operations.

### 6.5. Elliptic Curve Domain Parameters

Apart from the curve parameters a and b, there are other parameters that must be agreed by both parties involved in secured and trusted communication using ECC. These are domain

International Journal of Computer Science & Information Technology (IJCSIT), Vol 3, No 3, June 2011

parameters. The domain parameters for prime fields and binary fields are described below. Generally the protocols implementing the ECC specify the domain parameters to be used.

### 6.5.1. Domain Parameters for Elliptic Curve over Field F(p)

The domain parameters for Elliptic curve over F(p) are p, a, b, G, n and h. p is the prime number defined for finite field Fp . a and b are the parameters defining the curve $y2 = x3 + ax + b \pmod{p}$. G is the generator point (xG, yG), a point on the elliptic curve chosen for cryptographic operations. n is the order of the elliptic curve. The scalar for point multiplication is chosen as a number between 0 and n-1. h is the cofactor where h = #E(Fp)/n. #E(Fp) is the number of points on an elliptic curve.

### 6.5.2 Domain Parameters for Elliptic Curve over F(2m)

The domain parameters for elliptic curve over F(2m) are m, f(x), a, b, G, n and h. m is an integer defined for finite field F(2m). The elements of the finite field F(2m) are integers of length at most m bits. f(x) is the irreducible polynomial of degree m used for elliptic curve operations. a and b are the parameters defining the curve $y2 + xy = x3 + ax2 + b$. G is the generator point (xG, yG), is a point on the elliptic curve chosen for cryptographic operations. n is the order of the elliptic curve. The scalar for point multiplication is chosen as a number between 0 and n-1. h is the cofactor where h = #E($F_2^m$)/n. #E($F_2^m$) is the number of points on an elliptic curve.

### 6.7. Basic ECC Protocol Algorithms

A generalized overview of EC cryptographic algorithms for key agreement and digital signature are explained below.

### 6.7.1. Elliptic Curve Digital Signature Algorithm – ECDSA

Signature algorithm is used for authenticating a device or a message sent by the device. For example consider two devices A and B. To authenticate a message sent by A, the device A signs the message using its private key. The device A sends the message and the signature to the device B. This signature can be verified only by using the public key of device A. Since the device B knows A's public key, it can verify whether the message is indeed send by A or not. ECDSA is a variant of the Digital Signature Algorithm (DSA) that operates on elliptic curve groups. For sending a signed message from A to B, both have to agree up on Elliptic Curve domain parameters. The domain parameters are defined in section Elliptic Curve Domain parameters. Sender 'A' have a key pair consisting of a private key dA (a randomly selected integer less than n, where n is the order of the curve, an elliptic curve domain parameter) and a public key QA = dA * G (G is the generator point, an elliptic curve domain parameter).

### 6.7.2. Elliptic Curve Diffie Hellman Key Exchange – ECDH

ECDH is a key agreement protocol that allows two parties to establish a shared secret key that can be used for private key algorithms. Both parties exchange some public information to each other. Using this public data and their own private data these parties calculates the shared secret. Any third party, who doesn't have access to the private details of each device, will not be able to calculate the shared secret from the available public information. For generating a shared secret between A and B using ECDH, both have to agree up on Elliptic Curve domain parameters. The domain parameters are defined in section Elliptic Curve Domain parameters. Both end have a key pair consisting of a private key d (a randomly selected integer less than n, where n is the order of the curve, an elliptic curve domain parameter) and a public key Q = d * G (G is the generator point, an elliptic curve domain parameter). Let (dA, QA) be the private key - public key pair of A and (dB, QB) be the private key - public key pair of B.

94

The end A computes K = (xK, yK) = dA * QB

The end B computes L = (xL, yL) = dB * QA

Since dAQB = dAdBG = dBdAG = dBQA.

Therefore K = L and hence xK = xL

Hence the shared secret is xK

Since it is practically impossible to find the private key dA or dB from the public key K or L, it's not possible to obtain the shared secret for a third party.

## 7. VARIOUS IMPLEMENTATIONS OF ECC

After selecting a suitable set of algorithms, ECC is implemented on either hardware or software platform. Hardware implementation is considered as the most suitable option looking into the large key sizes and slow speed of point multiplication, when come together with limited resources of embedded platform. However, for application specific systems, embedding a separate piece of hardware for cryptography increases the manufacturing cost drastically. Also, various recent researches show that a careful selection of efficient algorithms and proper ECC parameters, ECC can be successfully implemented on software platforms. But, software implementations cannot match up to the speed of hardware implementations. Therefore research in the field of Elliptic curve cryptography has been propagated into both directions of hardware and software solutions.

A hardware implementation is basically focused on the processor design issues. To design any processor we have to consider flow of data into its input, output registers and various ALU units, called processor data path. Also flow of control information in the form of op-code and related micro-operations has to be considered called control logic. It is implemented in different ways like, extending instruction set of the processor or enhancing design of multiplier inside ALU. Following are some of the design considerations -

- Enhancing the speed of multiplication.
- Providing suitable options for very large data size to be implemented on data path of limited data size processor.
- Providing suitable options to implement various non standard number formats, used in cryptographic applications.
- Modifying the instruction set of domain specific processors for ECC operations
- Performing ECC on server, multi-core systems, media processors etc.

Below is the summary of some of the papers in this regard -

[23] and [24] have presented a pyramid hierarchy suggesting implementation of ECC for embedded systems in desired form. These papers discuss that the lower layers that represent choice of hardware design and processor group selection are more general and can be shared with other application specific pyramids. These layers represent FPGA, flash and ASIC design options and instruction accurate/cycle accurate models of RTL, strongArm etc.

The upper layers of hierarchy represent the number theory, security algorithms and security protocol architecture specific to cryptographic applications and rarely shared with other application.

Various designs of cryptographic processors are presented in [23], [25], [26], [27].

In [23] an ECC processor is presented for key generation and key agreement. The data path is bit sliced and n bit wide for data size of $GF(2^n)$. It is implemented using feedback loop in place of pipelining as feedback is needed for repeated point doubling and adding of ECC. The paper presents three-tier architecture of the processor. The lower layer implements the data path. The middle layer implements point doubling, addition and subtraction in two-way handshake. It means the processor can issue as well receive ECC keys. The top layers combine these operations in ECC multiplications as shown in Fig. 4. The instruction set of this processor includes various Elliptic Curve related operations. This processor can be implemented easily in a system and can run at an unrelated clock.

The processor design presented in [25] is based on the idea that general-purpose processors support multiplication operations very well. A design extension of the general-purpose processor is presented to cope with large data size of cryptographic applications. This processor supports famous RSA and Deffie- Hellman along with ECC. Hardware acceleration is considered on server machines. This processor supports RSA for GF(p) and arbitrary elliptic curves over fields GF(2n). Double and add and Montgomery scalar point multiplication operations are used for RSA and ECC. General-purpose processors typically have a data path width of 8, 16, 32, or 64 bits and operate on operands that have sizes equal to the data path widths. Thus, the long operands of the RSA and ECC algorithms need to be broken up into smaller words, and the arithmetic operations addition, subtraction, and multiplication need to be implemented as multiple-precision operations. As paper focuses on server applications, 64-bit architecture was considered. To adhere to a general-purpose data path, the amount and scope of modifications is limited; a number of performance optimizations were omitted. In particular, hardware division and optimized squarer for GF(2n) is not provided. The machine has two processors, one arithmetic processor and one control processor. VLIW(very large instruction word) architecture is used to keep both the processors busy at same time. For example, when a multiple-precision multiplication is performed, the arithmetic processor executes multiply-accumulate instructions while the control processor, in parallel, executes loop control instructions. This way, the critical path of the program execution is determined by the arithmetic instructions and the control instructions do not add any execution time. The final reduced instruction set for public key cryptography includes five arithmetic and four control instructions, a total of nine instructions only. The performance analysis shows clear performance advantage of ECC over RSA.

A scalable crypto co processor design is presented in [26]. This paper shows a word based (16 bit) co processor for RSA and ECC cryptography. It works exclusively for performing word based (16 bit) multiplication using Montgomery modular multiplication. It can multiply 2048 bits for RSA and 512 bits for elliptic curve cryptography. The design of the processor is such that we can use extra memory if data size has to be increased. The design of the processor includes I/O interface, a memory module, a dual field module, Crypto controller and RSA/ECC controller.

The embedded multicore systems for ECC are considered in [24]. Multicore processors for embedded systems are now getting increasing interest from the embedded system industry. Chip vendors like Intel, ARM, Fujitsu and Renesas have announced different multi-core processors for embedded systems. The advantages of using multicore processors in embedded systems can be simply summarized into two points. First, it gives the embedded developers more spaces to add new applications. Second, by using parallel computing we can reduce the clock frequency and voltage to achieve an energy efficient design. One major problem is data transfers between different cores. For multicore systems, software designers have to choose a good parallel computing strategy that fits the platform architecture. The above mentioned paper compares the performance of two parallel computing methods and tries to combine them efficiently.

Horizontal Parallelism is used to perform Montgomery modular multiple-precision operation (MMO) with all the cores in parallel. Vertical Parallelism is used to perform Different MMOs that have no data dependencies. In order to utilize all the cores efficiently, these horizontal parallelism and vertical parallelism are combined making a two-dimensional parallelism. Two dimensional parallelism starts from the schedule generated by vertical parallelism method, and deploy the horizontal parallelism method to perform some of the modular operations so that all cores in the system can be utilized.

Other papers like [28], [29], [30] also present various suggestions for processor designs. Most of the suggestions are focussed on how to enhance the speed of point multiplications on various types of processors. Modification in the design of multiplier, extending the instruction set of the processors for ECC specific operations, maximizing the utilization available processor resources are various design suggestions provided. Some of the papers are also available for the processor design on specific values of n in Zn [29], [30], [31]. Another design consideration is with respect to large data size of crypto – applications Vs. limited data type lengths available for embedded systems. Multiprecision arithmetic operations, i.e. performing arithmetic operations by dividing n bit data into blocks of k bits where each block is of size n/k and accumulating the final result is the common method for cryptographic related data implemented with various parallel processing techniques.

If cryptographic algorithms are implemented on hardware, they show higher performance in terms of speed. Also, they are more secure because they cannot be easily modified or read by an outside attacker. Still price per piece and power consumption are the main disadvantages. Such implementations are also inflexible in terms of ECC parameters making such systems vulnerable to attackers. We cannot switch between multiple schemes easily and if it is needed, it results into higher development cost.

Software implementations of ECC have the advantage of interoperability; we can enhance security while switching between different ECC schemes. Therefore ECC implementation using software is also seriously considered by the researchers. Various papers in this regard are [32] [33] [34]

Research in this category of ECC is mainly related to present various combinations of speed optimized algorithms that will reduce the requirement of separate crypto co processor. Also different mathematical techniques are considered to enhance the speed and security of ECC. A patent is assigned to sun Microsystems Inc for "Generic implementations of elliptic curve cryptography using partial reduction" [32]. A reduction operation is utilized in an arithmetic operation on two binary polynomials X(t) and Y(t) over GF(2) were an irreducible polynomial $M_m(t) = t^m + a_{m-1}t^{m-1} + a_{m-2}t^{m-2} + .... + a_1 t + a_0$, where the coefficients are equal to either 1 or 0, and m is a field degree. The reduction operation includes partially reducing a result of the arithmetic operation on the two binary polynomials to produce a congruent polynomial of degree less than a chosen integer n, with m ≠ n. The partial reduction includes using a polynomial $M', M' = (M_m(t) - t^m)^* t^{n-m}$, or a polynomial $M'', M'' = (M_m(t)^* t^{n-m})$ as part of reducing the result to the degree less than n and greater than or equal to m. The integer n can be the data path width of an arithmetic unit performing the arithmetic operation, a multiple of a digit size of a multiplier performing the arithmetic operation, a word size of a storage location, such as a register, or a maximum operand size of a functional unit in which the arithmetic operation is performed.

Hasegawa et al. implemented ECDSA signature generation and verification on a 10MHz M16C microcomputer [35]. The implementation requires 4KB of code space and uses a 160 bit field prime p = 65112 *2144−1 chosen to accommodate the 16-bit processor architecture. Signatures can be generated in 150ms and verified in 630ms. Based on the ECC integer library, the authors also estimate 10s for RSA-1024 signature generation and 400ms for verification using e = 216 +1.

In [33] it is discussed how ECC is efficient over RSA when implemented on an 8-bit processor. Following key steps were taken to accelerate performance of ECC on 8-bit processors.

Only prime integer fields are considered. Mixed coordinate systems using a combination of modified Jacobian and affine coordinates is used as they have been proved to offer the best performance [21].

They have used NIST and SECG specified set of elliptic curves as they have verified security properties and they allow for significant performance optimizations.

Non-adjacent forms (NAFs) are used. It is a method of recoding the scalar k in a point multiplication kP in order to reduce the number of nonzero bits and thus the number of point additions. This is accomplished by using digits that can be 0, 1 or -1. For example, 15P = (1 1 1 1)2P can be represented as 15P = (1 0 0 0 -1)2P.

As performing multiple precision multiplications on small processors involve significant amount of data transport to and from memory due to limited register space. Therefore to optimize ECC modular multiplication authors have considered cutting down the memory access per multiplication. Three multiplication strategies row wise multiplication, column wise multiplication and hybrid multiplication are considered.

Row-wise implementation requires n + 2 registers and performs n2 + 3n memory accesses. Column-wise multiplication requires $4 + \lceil \log_2(n)/k \rceil$ registers, the fewest number of all three algorithms. The number of registers grows only negligibly with the increase of the operand size n.

Hybrid multiplication aims at optimizing for both the number of registers and the number of memory accesses. The column-wise strategy is employed as the "outer algorithm" and the row-wise strategy as the "inner algorithm". That is, hybrid multiplication computes columns that consist of rows of partial products. Register usage and memory accesses depend on the number of partial products per row (or column width) d. It is suggested that d can be chosen according to the targeted processor; larger values of d require fewer memory operations, but more registers to store operands and to accumulate the result. To optimize the algorithm performance for r available registers, d should be chosen such that

$$d = \max\{i \mid 1 \leq i \leq n, r \geq 3i + 1 + \lceil \log_2(n/i)/k \rceil\}.$$

8 bit platforms chosen for implementations were Chipcon CC1010 8 bit 14.7456MHz microcontroller which implements Intel 8051 instruction set and ATmega128 8 bit microcontroller based on AVR architecture. It can be operated at frequencies upto 16 MHz.

A similar research presented in [36] shows that elliptic curve cryptography not only makes public-key cryptography feasible on highly constrained embedded devices like 'Motes', it allows one to create a complete secure web server stack that runs efficiently within very tight resource constraints. The small-footprint HTTPS stack, nicknamed Sizzle (Small SSL), has been implemented on multiple generations of the Berkeley/Crossbow motes where it runs in less than 4KB of RAM, completes a full SSL handshake in 1 second (session reuse takes 0.5 seconds) and transfers 1KB of application data over SSL in 0.4 seconds. The authors have claimed that Sizzle is the world's smallest secure web server and can be embedded inside home appliances, personal medical devices, etc., allowing them to be monitored and controlled remotely via a web browser without sacrificing end-to-end security.

A comparative performance evaluation for Pocket PC and wireless sensors is given in [37] to study the computational ability to process cryptographic functions, such as point multiplication, Pairings, AES, and hash functions. They have shown that current Pocket PC level devices are capable to process computational intensive cryptographic functions, such as Parings. However, purely software cryptographic solutions require long time to process cryptographic algorithms and special optimization methods must be used to improve the computation performance. They have used HP iPAQ hx4700 Pocket PC that runs Microsoft Windows Mobile 2003 operating system. The processor in the Pocket PC is a 32-bit processor. The application for the Pocket PC was developed in Microsoft Visual Studio 2005 as a smart device project using a machine running Windows XP. A software library from Shamus Software Ltd called Miracl is used for implementing pairing, hash and encryption algorithms in Pocket PC.

Micaz 2400 wireless sensor running TinyOS was used as a testing environment to calculate the time taken for primitive field operations in pairing. The central idea in Pairing Based Cryptography is the construction of a bilinear mapping between two useful cryptographic groups (ECC points group and OEF), which transfer the properties of one group to another. The sensor is connected to a MIB 510 base station, which is in turn connected to the personal computer via an USB port. The code was developed in TinyOS operating system using NESC language. The software package TinyECC is used, developed by North Carolina State University research group for implementing primitive field operation in pairing.

The authors have concluded that purely software-based solutions are suitable for Pocket PC; whereas wireless sensors have the difficulty to compute computational intensive operations, such as Parings. The have also aimed at building an efficient pairing implementation in sensors by

- applying efficient Pairings algorithms for sensors
- identifying the most efficient curve and system parameters for Parings in sensors
- specially optimizing the field operations for sensors.

A recent Paper by Johan Dams [38] is focused on performance comparison of ECC on medium sized embedded systems. The paper is aimed to design the system to which is;

- Portable: the system should be easy to port to different hardware platforms
- Fast: optimizations should be made where possible, but assembly code should be limited as it hampers first requirement.
- Efficient: Use as little system resources as possible, yet don't overuse assembly language as this hinders portability.

Following systems were considered for implementation. All systems are running Linux, with either GNU Libc or uClibc.

- Nokia N800, TI OMAP 2420 clocked at 330MHz (GNU Libc)
- Freescale MPC5200B clocked at 400MHz (GNU Libc)
- Renesas SH7203 clocked at 200MHz (uClibc)
- Freescale ColdFire MCF54455 clocked at 266MHz (GNU Libc)
- Freescale ColdFire MCF52277 clocked at 160MHz (uClibc)

The performance is compared for time needed by each CPU to perform field operations, point multiplications, encryption/decryption, signature generation/ verification and key generation.

Finally, let us take a review of Various ECC standards as proposed by NIST and other agencies.

**American National Standards Institute (ANSI)** The ANSI X9F subcommittee of the ANSI X9 committee develops information security standards for the financial services industry. Two elliptic curve standards have been completed: ANSI X9.62 which specifies the ECDSA (§4.4.1), and ANSI X9.63 which specifies numerous elliptic curve key agreement and key transport protocols including STS (§4.6.1), ECMQV (§4.6.2), and ECIES (§4.5.1). The objective of these standards is to achieve a high degree of security and interoperability. The underlying finite field is restricted to being a prime field Fp or a binary field F(2m). The elements of F2m may be represented using a polynomial basis or a normal basis over F2. If a polynomial basis is desired, then the reduction polynomial must be an irreducible trinomial, if one exists, and an irreducible pentanomial otherwise. To facilitate interoperability, a specific reduction polynomial is recommended for each field F2m these polynomials of degree m, where $2 \leq m \leq 600$.

**National Institute of Standards and Technology (NIST)** NIST is a non-regulatory federal agency within the U.S. Commerce Department's Technology Administration. Included in its mission is the development of security-related Federal Information Processing Standards (FIPS) intended for use by U.S. federal government departments. The FIPS standards widely adopted and deployed around the world NIST special publication 800-57[39] indicates the value of f (the size of n, where n is the order of the base point G) for algorithms based on elliptic curve cryptography (ECC) that are specified for digital signatures in [ANSX9.62] and adopted in [FIPS186-3], and for key establishment as specified in [ANSX9.63] and [SP800-56]. The value of f is commonly considered to be the key size as per the table summarized below.

Table 2. NIST recommended key length

| Algorithm security lifetimes | ECC (e.g. ECDSA) |
| --- | --- |
| Through 2010 (min. of 80 bits of strength) | Min.: f=160 |
| Through 2030 (min. of 112 bits of strength) | Min.: f=224 |
| Beyond 2030 (min. of 128 bits of strength) | Min.: f=256 |

**Institute of Electrical and Electronics Engineers (IEEE)** The IEEE P1363 working group is developing a suite of standards for public-key cryptography. The scope of P1363 is very broad and includes schemes based on the intractability of integer factorization, discrete logarithm in finite fields, elliptic curve discrete logarithms, and lattice-based schemes. The 1363-2000 standard includes elliptic curve signature schemes (ECDSA and an elliptic curve analogue of a signature scheme due to Nyberg and Rueppel), and elliptic curve key agreement schemes (ECMQV and variants of elliptic curve Diffie-Hellman (ECDH)). It differs fundamentally from the ANSI standards and FIPS 186-2 in that there are no mandated minimum security requirements and there is an abundance of options. Its primary purpose, therefore, is to serve as a reference for specifications of a variety of cryptographic protocols from which other standards and applications can select. The 1363-2000 standard restricts the underlying finite field to be a prime field Fp or a binary field F2m. The P1363a draft standard is an addendum to 1363-2000. It contains specifications of ECIES and the Pintsov-Vanstone signature scheme providing message recovery, and allows for extension fields Fpm of odd characteristic including optimal extension fields.

**American National Standards Institute (ANSI)** The ANSI X9F subcommittee of the ANSI X9 committee develops information security standards for the financial services industry. Two elliptic curve standards have been completed: ANSI X9.62 which specifies the ECDSA (§4.4.1), and ANSI X9.63 which specifies numerous elliptic curve key agreement and key transport protocols including STS (§4.6.1), ECMQV (§4.6.2), and ECIES (§4.5.1). The objective of these standards is to achieve a high degree of security and interoperability. The underlying finite field is restricted to being a prime field Fp or a binary field F(2m). The elements of F2m may be represented using a polynomial basis or a normal basis over F2. If a polynomial basis is desired, then the reduction polynomial must be an irreducible trinomial, if one exists, and an irreducible pentanomial otherwise. To facilitate interoperability, a specific reduction polynomial is recommended for each field F2m these polynomials of degree m, where 2 ≤ m ≤ 600.

**International Organization for Standardization (ISO)** ISO and the International Electrotechnical Commission (IEC) jointly develop cryptographic standards within the SC 27 subcommittee. ISO/IEC 15946 is a suite of elliptic curve cryptographic standards that specifies signature schemes (including ECDSA and EC-KCDSA), key establishment schemes (including ECMQV and STS), and digital signature schemes providing message recovery. ISO/IEC 18033-2 provides detailed descriptions and security analyses of various public-key encryption schemes including ECIES-KEM and PSEC-KEM.

**Standards for Efficient Cryptography Group (SECG)** SECG is a consortium of companies formed to address potential interoperability problems with cryptographic standards. SEC 1 specifies ECDSA, ECIES, ECDH and ECMQV, and attempts to be compatible with all ANSI, NIST, IEEE and ISO/IEC elliptic curve standards. Some specific elliptic curves, including the 15 NIST elliptic curves, are listed in SEC 2.

## 8. CONCLUSIONS

Elliptic curve cryptography has been emerged as a vast field of interest for application specific security requirements. It has its roots into the number theory which was already used for cryptographic applications before ECC. The elliptic curve discrete logarithm problem makes ECC most efficient with smaller key size compared to earlier RSA algorithm. It is mostly considered for resource constrained devices. Research in the field of Elliptic Curve Cryptography has emerged in various directions to analyze its proper implementation on hardware as well as software platforms.

**Authors**

**Dr. Suresh C. Mehrotra**, F.N.A.Sc, FIETE , has received M.Sc.(Physics) from Allahabad university in 1970 & Ph.D from University of Texas (Austin) in 1975.He is recipient of UGC career award, Welch Foundation fellow 1972 (U.S.A), Dutch fellowship 1980 (Netherlands), NASA Fellowship 1981 (New York), Alexander Von Fellowship 1983 (W. Germany ) etc. He joined Physics Department of Dr. Babasaheb Ambedkar Marathwada University in 1979 as Reader and then became Professor in 1996. Since 1997, he has worked as a Professor in Department of Computer Science and Information Technology. He is actively engaged in research and had published more than hundred & fifty papers in International and National reputed journals .He is reviewer for several reference journals for India and abroad. His areas of interest are Simulation of Computer Networks, Speech Recognition, Character Recognition, Facial Expression Recognition, Human Computer Interaction, Microwaves, and Simulation of Liquids.

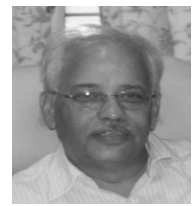

**Khan Rahat Afreen,** B.E (CSE), M.E, Persuing PhD Under the guidance of Dr. S.C. Mehrotra for the topic "Implementing Elliptic Curve Cryptography for Embedded Systems Using Hash Table to Achieve Large Key Size" Presented papers related to Cryptography in National Workshop on Cryptology-2006 held at Defense Institute of Advanced Technology, Pune, and in the International Conference on "Advances in Computer Vision and Information Technology – ACVIT-07 held at Dept. of Comp. Science & I.T. Dr. Babasaheb Ambedkar Marathwada University, Aurangabad. Currently working as Associate Professor and Incharge Principal at Tom Patrick Institute of Computer & Information Technology, Dr. Rafiq Zakaria Campus, Aurangabad.

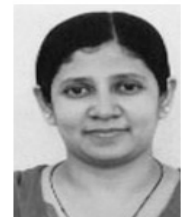